# Towards an extension of 1905 relativistic dynamics with a variable rest mass measuring potential energy

Rafael A. Valls Hidalgo-Gato [1]

[1] ICIMAF, Cuba, valls@icimaf.cu

*Abstract* — From a rigorous historic analysis of 1686 I. Newton and 1905 A. Einstein works where the last derived the universal mass-energy relationship, it is concluded that rest mass measures potential energy. From the same formula used to obtain that relation, it is derived the ratio Total Energy/Potential Energy is equal to the γ relativistic factor. It is derived a formula for the variation of a body rest mass with its position in a gravity field, explaining with it the behavior of an atomic clock. It is revised the bodies free fall in a gravitational field, finding that a constant total mass is equal to the gravitational mass, while the variable rest mass is equal to the inertial mass, maintaining all an identical behavior independent of their masses. A revision of the Eötvös experiment concludes that it is unable to detect the found difference between inertial and gravitational mass. Applying the extended 1905 relativistic dynamics to Mercury, its perihelion shift is determined; it is concluded with the convenience to continue its development, what can imply a revision of Physics since 1905 with important results in the unification of natural forces and other open problems.

## I. INTRODUCTION

In recently published papers, we had explained the behavior of an atomic clock in a gravitational field [1]-[2] and the Mercury's perihelion shift [3], employing exclusively the 1905 Relativity (1905R), considering 1905R only the first year of the denoted by A. Einstein in 1916 [4] as Special Relativity (SR), to distinguish it from his General Relativity (GR).

Taking into account that until now it was considered that the mentioned physical effects only could be explained using the GR, we had considered adequate in this paper to show the reached results in [1]-[2]-[3] following a different order, to emphasize what we considered the primary fact that made possible to explain such effects with only 1905R, that a body rest mass (RM) measures its potential energy (PE).

For the first effect, close related with the denoted GR gravitational red shift, an alternative explanation is given that does not appear at first view to be out from the GR scope; but the second effect is explained deriving from the 1905R that the inertial mass (IM) is *not* equal to the gravitational one (GM), but only to a part of it that results equal to the RM that measures the PE. As we will see in section III, the ratio GM/IM results equal to the 1905R relativistic factor today denoted γ, where the validity of the equations of the Newtonian mechanics is a definition requirement for what 1905 Einstein denotes in [5] as *stationary system*, basic concept of his new theory. The implications for the relativistic mechanics are emphasized in the present paper title.

## II. HOW DERIVES EINSTEIN IN 1905 THAT THE MASS MEASURES ENERGY

At 3 months from his first paper on Relativity [5], Einstein publishes another (very short, only 3 pages) where derives the universal relationship between mass and energy.

Einstein begins considering a stationary body (at rest) in a system $S_1$ with energy $E_0$, having the same body energy $H_0$ in another system $S_2$ in which it is moving with velocity *v*. Considers then that some part L of $E_0$ is emitted as light out from the body (in two halves with equal directions and opposite senses, such that the body continues at rest in $S_1$), applying to both systems the Principle of Energy Conservation according with his Principle of Relativity (the same laws in all systems). Using a formula derived in [5] that relates light energies in different systems, and some elemental algebraic operations, reaches to the following expression

$$H_0 - E_0 - (H_1 - E_1) = L \{[1/\sqrt{(1 - v^2/c^2)}] - 1\}, \qquad (1)$$

where sub-indexes 0 and 1 indicate before and after the emission of light respectively. Results crucial to interpret the text that follows in the rigorous 1905 context:

"The two differences of the form H − E occurring in this expression have simple physical significations. H and E are energy values of the same body referred to two different systems of co-ordinates ($S_2$ and $S_1$ respectively) which are in motion relatively to each other, the body being at rest in one of the two systems ($S_1$). Thus it is clear that the difference H − E can differ from the kinetic energy K of the body, with respect to the other system ($S_2$), only by an additive constant C, which depends on the choice of the arbitrary additive constants of the energies H and E. Thus we may place

$$H_0 - E_0 = K_0 + C, \qquad (2)$$
$$H_1 - E_1 = K_1 + C, \qquad (3)$$

since C does not change during the emission of light. So we have

$$K_0 - K_1 = L \{[1/\sqrt{(1 - v^2/c^2)}] - 1\}". \quad (4)$$

Putting (2) and (3) as $H = K + (E + C)$, it results evident that in the 1905 context, the total energy H can not be any other thing that the kinetic energy K plus the *potential energy* E with its characteristic arbitrary additive constant C. The explicit handling that makes Einstein with the arbitrary additive constants that characterize the potential energies does not leave place to any other interpretation. Such as he declares at the beginning of his paper, Einstein makes use of the Principle of Energy Conservation, in the unique way compatible with the historic context: Total Energy (H) = Kinetic Energy (K) + Potential Energy (E). Only in that way results clear (as Einstein says) the introduction of (2) and (3).

Once Einstein concludes that the mass of a body is a measure of its energy-content (without excluding any type of it), corresponds to its rest mass to measure its rest energy $E_0$ (that is precisely from where light energy L is taken, diminishing the rest mass in $L/c^2$). But all we know that the development of relativistic Physics followed another road: the body rest mass was considered an intrinsic constant, without any relation with its potential energy.

Have no sense at all to interpret (in 1905) that the $E_0$ is a new type of energy (without any relation with the potential one) measured by the rest mass. How can we suppose that the mass measures energy, if we are precisely starting to analyze the paper where for first time that conclusion is reached? Before being measured by a mass, $E_0$ *must* be *before* an energy recognized in 1905 to which the Principle of Energy Conservation can be applied *afterward*, and if the body is at rest, it is clear that it can not be kinetic, *not* resting other alternative than the potential energy (of *all* types that could be present, known or not, including the gravitational potential energy). What other energy type known in 1905 (not being the potential) could be transformed in the kinetic energy of the light emitted following the Principle of Conservation? And even supposing that it existed, what reason could exist to exclude the potential energy, being well known in the epoch Physics its ability to convert in kinetics and vice versa, that is precisely the expression form in Mechanics of the Principle of Energy Conservation that applies Einstein?

In the literature we find a large debate (that reaches our days) [7]-[8]-[9]-[10]-[11] about the content of [6], with different interpretations and often contradictory. The disagreement is lumped in the validity grade of the original derivation, being accepted the universal mass-energy equivalence as a physical fact with huge experimental evidence. The majority of the interpretations mentioned are realized much after 1905. We will not analyze them, considering out of the historic context specified for the present paper.

In 1965 Leon Brillouin [12]-[13] did intent correct the handling of the potential energy in Special Relativity, attributing mass to the field potential energy, but a mass different to the body rest mass (that continued considering, as everybody, an intrinsic constant independent from potential energy, practice that is maintained until today). In our interpretation (the unique one that we considered correct in the 1905 historic context), we coincide with Brillouin in that the field has mass that measures its potential energy, but (different from him) we considered that this mass is the *same* rest mass that we attribute to the body that has associated the field.

### III. RELATION BETWEEN THE TOTAL ENERGY AND THE POTENTIAL ENERGY

Let us see now what happens if in the same formula (4) from where Einstein derives the universal relationship between mass and energy, we considered the rest mass playing the role that corresponds to it as a measure of the potential energy.

If L is any part of $E_0$, let us see the case $L = E_0$, i.e., that all the energy of the rest body pass to be emitted light. In this case the original body disappears, not having as a result any kinetic energy after the emission, i.e., $K_1 = 0$. Denoting $K_0 = KE$ the kinetic energy and $L = E_0 = EP$ the potential energy, both in the system $S_2$, we obtain then

$$KE = PE \{[1/\sqrt{(1 - v^2/c^2)}] - 1\} = (\gamma - 1) PE, \quad (5)$$

where γ is the known 1905R relativistic factor. And as Total Energy (TE) = Kinetic Energy (KE) + Potential Energy (PE), we obtain then after a simple algebraic transform that

$$[\text{Total Energy (TE)}] / [\text{Potential Energy (PE)}] = \gamma. \quad (6)$$

Taking into account that γ is a (scalar) function of a body (vector) velocity *v*, the previous expression revels us a very general (universal) relationship among Total Energy, Potential Energy, Kinetic Energy, velocity and speed for any body.

Note that if we interpret PE as the "proper energy" of the Special Relativity (SR), we obtain the same SR formula that relates the increase in total energy with an increase in the speed. Instead of a constant PE with a variable TE, we have (in the two cases we considered later) a constant TE (associated to the Principle of Energy Conservation) with a variable PE (measured by a variable rest mass).

The previous coincidence suggest that, instead of substituting the SR relativist dynamics valid for a free body, what we really are doing is extending its application to a bound body, a fact taking into account when assigning a title to this paper.

As the mass measures the energy (no matter if it is partial or total), the ratio between the energies found in (6) is also equal to the ratio between the corresponding masses, obtaining then

$$[\text{Total Mass (TM)}] / [\text{Rest Mass (RM)}] = \gamma, \quad (7)$$

## IV. VARIATION OF THE REST MASS IN A GRAVITATIONAL FIELD

To fix the historic context in which the following derivation is accomplished, we consider appropriate to refer the beginning of §1 in [5]:

"Let us take a system of co-ordinates in which the equations of Newtonian mechanics hold good. In order to render our presentation more precise and to distinguish this system of co-ordinates verbally from others which will be introduced hereafter, we call it the *stationary system*.
If a material point is at rest relatively to this system of co-ordinates, its position can be defined relatively thereto by the employment of rigid standards of measurement and the methods of Euclidean geometry, and can be expressed in Cartesian co-ordinates."

In what follows we use the Newtonian concept of gravitational potential and the Euclidean geometry, with polar coordinates for the central gravitational field that we consider. Following 1905 Einstein, we consider any centre of mass Newtonian system (corresponding to any determined body set modeled by material points) as a *stationary system* in which are valid the equations of Newtonian mechanics.

Let be two bodies modeled by the material points M and m (one with a great mass M and the other with a small m<<M). The centre of mass of the corresponding *stationary system* coincides then practically with the centre of mass of M (here M can be for example the Earth and m an electron, or M the Sun and m Mercury). We denote as $r$ the position vector of m, with (scalar) distance r from M.

As more far away is m from M, so greater will be its Potential Energy (PE), measured by its rest mass $m_0(r)$. If we denote $m_{0m}$ its maximal value at infinite, we have then that

$$PE(r) = m_0(r)c^2 = m_{0m}c^2 - (GM/r) m_0(r) \quad (8)$$

where G is the Newtonian gravitational constant, c the vacuum light speed and (GM/r) the gravitational potential associated to M (PE by unit of the $m_0$ of the body situated at $r$). As r tends to infinite, $m_0$ tends to $m_{0m}$ and PE to $m_{0m}c^2$ as corresponds. From (8) we obtain easily

$$m_0(r) = m_{0m} / (1+GM/rc^2). \quad (9)$$

As we see, the arbitrary additive constant characteristic of potential energy disappears in 1905R, appearing a PE zero point in r = 0 that is not arbitrary, but a consequence of the rest mass measuring potential energy. If M is the Earth and m an electron, $m_{0m}$ is the ordinary rest mass of a *free* electron (the today considered intrinsic constant).

## V. BEHAVIOR OF AN ATOMIC CLOCK IN A GRAVITATIONAL FIELD

Since 1913 (N. Bohr H model) [14], it is known that the frequency emitted by an atom is proportional to the rest mass (RM) of the electron that changes its energy state. In the Einstein's General Relativity (GR), the RM is an intrinsic attribute (constant) of the electron, being explained the change in frequency (inverse of time) for the curved space-time provoked by the presence of the mass-energy M.

In 1905R, taking as reference the maximal frequency f(∞) of an atomic clock at infinite, we can multiply by the factor of 1905R

$$f(r) / f(\infty) = 1/[1+(GM/rc^2)] \quad (10)$$

that we take from (9), obtaining the frequency f(r) at any position $r$. The corresponding GR factor is known (see for example [15]) to be

$$f(r) / f(\infty) = \sqrt{[1-(2GM/rc^2)]}. \quad (11)$$

The frequency change that predicts 1905R is very near to the RG one in all the range of practical r values ($GM/rc^2 <<1$), in real experiments like the Pound&Rebka [16] or in the continuous operation of the Global Positioning System (GPS) [17] of our days.

In R1905 the r can takes any value in the interval from 0 to infinite, while in GR can not do it for $r<2GM/c^2$ where the factor takes an imaginary value.

We want to emphasize the absent in 1905R of the singularity that appears in GR. Remains open the problem to determine if this absence implies a 1905R limitation that does not permit it to address a black hole, or by the contrary implies a theoretical evidence of its not existence with the singular attributes that predicts the GR.

## VI. BODIES FREE FALL IN 1905R

For the same two bodies M and m of section IV, we consider now the small m on free falling in the gravitational field of the great M. The general case would be m orbiting M, what as a degenerate particular case would be m falling from the height $r_m$ starting from rest. Applying the equations of Newtonian mechanics (corresponding to the universal gravitational law and the second law of mechanics), we reach to the following expression that gives us the acceleration a of the body m

$$a = F / m_i = (GM/r^2) (m_g / m_i), \quad (12)$$

where F is the gravitational force between both bodies, $m_g$ is the gravitational mass of m and $m_i$ its inertial mass. We know besides that during all the process the total mass m is maintained constant, taking place a transformation process of the potential energy in kinetic one or vice versa, measuring the always constant m the total mass, and the variable rest mass $m_0$ (function of the position *r*) the potential energy.

Let us put in explicit form the ratio ($m / m_0$) of (7) as a function of the speed (scalar) v of the body that moves

$$[m / m_0(v)] = 1 / \sqrt{(1 - v^2/c^2)} = \gamma. \qquad (13)$$

The expression (13) revels us clearly that the speed v of the body m in each instant, and then *also* its acceleration a (and all the other derivatives of superior order with respect to the time), stay completely determined by the *fraction* ($m_0 / m$) (inverse of γ) of the constant total mass m that represents the *variable* rest mass $m_0$ in such instant.

As in the process that we consider the total mass m remains always constant, this means that the variable rest mass $m_0$ results being the one determining the acceleration a of the body m in each instant when the force F is applied to it, i.e., the *rest mass* $m_0$ is behaving as the *inertial mass* $m_i$ of the body m.

Seeing (12) and (13), we realize that the ratio ($m_g / m_i$) that appears in (12) results being equal to the ratio ($m / m_0$) of (13) if besides of the equality ($m_i = m_0$) we consider the equality ($m_g = m$). This last equality results very reasonable, because remaining m always constant in all the process (as a consequence of the Principle of Energy Conservation), there is no reason at all for not being the body gravitational mass $m_g$ equal to its constant total mass m, as it is in the Newtonian mechanics, and remembering that the validity of its equations is a defining requirement of what in 1905R is denoted as *stationary system*, precisely the context in which we are revising the bodies free fall.

We knew already since Galileo that all the bodies fall with the same velocity and acceleration in each instant, no matter how different its total mass m can be. We thought that the unique possible cause were the equality between gravitational mass $m_g$ and inertial one $m_i$. Now we know that in 1905R we have

$$(m_g = m), \qquad (14)$$

$$(m_i = m_0), \qquad (15)$$

$$(m_g / m_i) = (m / m_0) = 1 / \sqrt{(1 - v^2/c^2)} = \gamma, \qquad (16)$$

being ($m_g = m_i$) only for a stationary body m.

## VII. REVISION OF EÖTVÖS EXPERIMENT IN 1905R

The Eötvös experiment [18] (original design about 1885) consists in a torsion balance where two bodies of different masses ($m_1 < m_2$) are put, hanging from a thin fiber placed in such a way that the lengths of the balance arms are inverse proportional to the masses, assuring that the torques of the gravitational forces (proportional to the respective gravitational masses) that the Earth applies to the bodies result neutralized mutually.

It is accomplished in any place of the Earth's surface where its rotation with lineal velocity v determine the existence of inertial forces (centrifugal) $F_1$ and $F_2$ (proportional to the respective inertial masses). It is reasoned that if gravitational masses are equal to the inertial ones, the torques of $F_1$ and $F_2$ remain also neutralized. As being not observed (with the great characteristic accuracy of the device) any spin, it is concluded as an experimental fact the equality of the gravitational mass with the inertial one.

Until our days nobody had put in doubt the validity of the experiment. However, in the previous section we derived from 1905R that the ratio ($m_g / m_i$) is equal for all the bodies, being ($m_g = m_i$) only in the case where the body is stationary. This means a diminution for the inertial forces if we compare them with the gravitational forces; but as this diminution is in a equal proportion for all the bodies, the corresponding torques result balanced as before even being in this case ($m_g > m_i$).

The Eötvös experiment is then *unable* to detect the difference between $m_g$ and $m_i$ when the ratio ($m_g / m_i$) is function of only the body's velocity, what is equal for all the bodies with the velocity v of any point of the Earth's surface where the experiment is executed. The inertial mass $m_i$ not only *can* be the rest mass $m_0$ as a part of the total mass m, but it *must* be in order to be coherence between the theoretical result derived from 1905R and the many practical results of the Eötvös experiment before and after 1905. In other words, that the Eötvös experiment not only does *not* contradict the difference between $m_g$ and $m_i$ derived from 1905R, but that *confirm* it.

## VIII. RELATION BETWEEN POSITION AND VELOCITY IN 1905R

In (9) we find the factor

$$m_0(r) / m_{0m} = 1 / (1 + GM/rc^2) \qquad (17)$$

what gives us how vary the rest mass of a small body m with its position *r* in the gravitational field of a great one M, taking as the reference its maximal value $m_{0m}$ at infinite. In a similar way, for the same case we find in (13) the factor

$$m_0(v) / m = \sqrt{(1 - v^2/c^2)} \qquad (18)$$

what gives us how vary the rest mass of the same small body m with its velocity *v*, but taking now as the reference its total mass m that remains constant, even if its position can vary in the gravitational field of the great body M within determined limits (aphelion and perihelion of m orbiting M, or maximal height $r_m$ since where falls m starting from the rest (v=0)). We are interested now in determine the ratio ($m_{0m} / m$) of the reference constants $m_{0m}$ and m. As m is the value of $m_0$ for r = $r_m$ (total energy equal to potential energy when v=0), from (9) we obtain then

$$m = m_0(r_m) = m_{0m} / (1 + GM/r_m c^2). \qquad (19)$$

From (9) and (13) we derive

$$m_{0m} / m = (1+GM/rc^2) \sqrt{(1 – v^2/c^2)}. \quad (20)$$

The ratio ($m_{0m} / m$) can be determined then with (20) from known values of r and v in *any* point of the trajectory of m. This theoretical prediction finds experimental support in the following section.

The constant character of the ratio ($m_{0m} / m$) makes that (20) established a mutual dependent relationship between position r and velocity v, that makes that known one of them, the other left then completely determined.

That position and velocity are determined mutually is nothing new for astronomers. What is possibly new is the fact that this relation is derived from 1905R, applicable inclusive to the Mercury's perihelion shift, as we see in the following section.

### IX. DETERMINING MERCURY'S PERIHELION SHIFT

Considering M the Sun's mass and m the Mercury's one, we are going to determine the ratio ($m_{0m} / m$) employing (20). We use astronomical data taken from [19].

We choose as points of Mercury's orbit its aphelion and perihelion, obtaining for both the same result (with 12 significant ciphers)

$$m_{0m} / m = 1,00000001275 \quad (21)$$

confirming with independent real data experimentally measured (actualized recently), the theoretical prediction of the previous section.

Taking into account that the two factors multiplied in (20) are the result of a theoretical derivation that has as the starting point the rest mass measuring potential energy, without any apparent previous link with the Newtonian mechanics that employ astronomers, results in extreme significant the match reached.

The value of m that appears in [19] is $0,3302 \times 10^{24}$ kg with 4 significant ciphers, what compared with the 7 zeros that appear in (21) indicates us that the best value that we can take for $m_{0m}$ is the same of m. In what follows we consider then with great security that

$$m_{0m} = m. \quad (22)$$

From (9) and (22) we obtain

$$m / m_0 = (1+GM/rc^2), \quad (23)$$

and from (16) and (23)

$$m_g / m_i = (1+GM/rc^2). \quad (24)$$

From (12) and (24) we obtain for the acceleration of Mercury

$$a = (GM/r^2) (1+GM/rc^2). \quad (25)$$

It is known that

$$(GM/r) = v_L^2, \quad (26)$$

where $v_L$ is the denoted lateral velocity of the planet, velocity component orthogonal to the vector position *r* (see for example [20]). Finally, from (26) and (25) we obtain

$$a = (GM/r^2)[1+(v_L^2/c^2)], \quad (27)$$

what results being *exactly* the same expression reached in [20] when Mercury's perihelion shift is determined from General Relativity (GR), sharing then 1905R and GR the *same* prediction.

### X. 1915 EINSTEIN VS. 1905 EINSTEIN

How it is possible that in 1905R, from which we recently derived that the gravitational mass is *not* equal to the inertial one, we reach to the *same* prediction that makes the 1915 GR starting from the Principle of Equivalence, what implies that the gravitational mass *is* equal to the inertial one?

We can find an adequate answer if we put our attention on the nature of Science in general and the way it developed. The man interacting with Nature creates theories (models) through the ones it is known each time better in an infinite process. An essential part of this process is the confirmation of theories through the experiments, what decide the acceptation grade or rejection of them, as also their modifications or creation of new ones.

In the case that occupy us, if the 1905R is able to explain in the future all (or at least the great majority) of the effects that explain RG, we have no doubt at all that it will end substituting it. The contradiction between two theories that compete between them is nothing new in Physics. It is sufficient to remember the wave and corpuscular light theories. More ever, the contradictions appears inclusive in the same theory, as is precisely the case today in the quantum-mechanics description of light and other corpuscles-waves.

Following the logic of Physics development, any theory can be submitted in any time to a revision that takes into account new facts, being them theoretical or experimental.

We consider opportune to mention that already Leon Brillouin at the end of [12] calls to revise the works of Sommefeld and Dirac establishing relations between Special Relativity and Quantum Mechanics. We can extend that revision more back away, in order to include at least the work of 1908 Minkowski [23] and even Einstein's ones between 1905 and 1908. In the following section we go really much more back away, revising 1686 Newton.

What we have in mind is to have the possibility to clear the way the intrinsic constant mass is introduced in Physics measuring a new and unknown until then energy, whose possible substitution by a variable rest mass that measures the potential energy is the principal topic of this paper.

The job is really great, from 1905 to today is more than a century, long period full of new developments in which had remained intact the constant nature of mass, without taking into account the huge experimental evidence that never stops to indicate that any expulsion of potential energy out of a system is always accompanied of the corresponding diminution of its rest mass (mass defect), always in the constant proportion ($c^2$) discovered by 1905 Einstein. But we all continue believing that the unique real energy that leaves the system left inside a mysterious phantom (*negative* binding energy) able to realize (no matter its without body nature) the important job to maintain united (stable) the system.

Taking into account the simple way (almost trivial) in what 1905R explains already effects traditionally associated to the GR (what needs to use a much more complicated mathematical and physical description), we not hide our sympathy for 1905 in this confrontation with 1915, as for equal results, the simple is the better, as when Copernicus (followed by Galileo and Newton) is confronting Ptolemy.

## XI. 1686 NEWTON GENERALIZING GALILEO'S PRINCIPLE OF RELATIVITY

In the following we show the last 3 Corollaries in [21], with the original Newton's text translated to English in1846.

Corollary IV:[The common centre of gravity of two or more bodies does not alter its state of motion or rest by the actions of the bodies among themselves, and therefore the common centre of gravity of all bodies acting upon each other (excluding outward actions and impediments) is either at rest, or moves uniformly in a right line.]
…..[For the progressive motion, whether of one single body, or of a whole system of bodies us always to be estimated from the motion of the centre of gravity.]

Corollary V: [The motions of bodies included in a given space are the same among themselves, whether that space is at rest, or moves uniformly forwards in a right line without any circular motion.] …..[A clear proof of which we have from the experiment of a ship; where all motions happen after the same manner whether the ship is at rest, or is carried uniformly forwards in a right line.]

**Corollary VI:** [*If bodies, any how moved among themselves, are urged in the direction of parallel lines by equal accelerative forces, they will all continue to move among themselves, after the same manner as if they had been urged by no such forces.*] …..[*For these forces acting equally (with respect to the quantities of the bodies to be moved), and in the direction of parallel lines, will (by Law II) move all the bodies equally (as to velocity), and therefore will never produce any change in the positions or motions of the bodies among themselves.*]

Corollary IV refers to the Newtonian centre of mass system corresponding to any body set, that can be considered a single whole body belonging to a body set of higher hierarchy (for example, the system Earth-Moon as part of the Solar system, or this last as part of the Galaxy).

The reference to Galileo's ship put in evidence that Corollary V corresponds to Galileo's Principle of Relativity, with acceleration zero (any uniform velocity) for all the bodies in the set. The corresponding centre of mass system can be used only to describe the movements of the bodies belonging to the set (only the bodies inside the ship, never the exterior ones).

Corollary VI is a generalization of Galileo's Principle of Relativity. The acceleration zero of Corollary V is generalized to *any variable acceleration*, with the requirement to be always the same for all the bodies in the set, that in this way always share a same velocity component. This requirement is for example well satisfied by the bodies of the Solar System that move as a single body in the Galaxy, or even by the System Earth-Moon that moves as part of the Solar System.

We had considered appropriate to refer 1686 Newton, not only for being the validity of his equations a definition requirement of the 1905 Einstein *stationary system,* but besides because his generalization of Galileo's Principle of Relativity could have relation with the reasons of 1915 Einstein to develop the new GR from SR, declaring since then this last *unable* to address gravitation, a thing that the results showed in this article contradicts. Only the SR *after* 1905 results unable to address adequately the gravitation. Do not forget us that Einstein in 1911 addresses the gravitational field [22], introducing the Principle of Equivalence that brings him to the development of GR. A century later we are showing that this road (or alternative), as we pointed already in [1]-[2]-[3], is not the unique possible.

## XII. CONCLUSIONS

In the present article had been showed in detail how the consideration of a rest mass measuring potential energy, fruit of a rigorous historical analysis of the way in which Einstein finds the equivalence mass-energy in 1905, lead us to completely unexpected results, finding explanation to effects only reached a decade later after the introduction of essential changes in the today denoted Special Relativity (SR) that lead to the development of General Relativity (GR).

All we know that GR, as the more advanced gravity theory, had not being yet possible to unite with the rest of natural forces, constituting today one of the most important research topics that remains open in Physics. Close related with it, we can consider the contradictions since their common origin had been existing until today with the two great theories of today Physics, General Relativity and Quantum Theory.

Taking into account now the new roads that open with a variable rest mass measuring the potential energy of *all* fields, known or by known, without any arbitrary additive constant, only rest to conclude with the convenience to continue for these new roads, with the hope to find a coherent development of Relativity and Quantum Mechanics since their common origin, what contributes to the solution of the great problems of today Physics.